\documentclass{webofc}
\usepackage[varg]{txfonts}   
\newcommand{\nf}{N_{\mathrm{f}}}
\newcommand{\mK}{m_\mathrm{K}}
\newcommand{\mpi}{m_\pi}
\newcommand{\mev}{\,\mathrm{MeV}}
\newcommand{\fm}{\,\mathrm{fm}}

\newcommand{\Nf}{N_{\mathrm{f}}}

\begin{document}
\title{Charmonium and exotics from lattice QCD}
%
%

\author{\firstname{Francesco} \lastname{Knechtli}\inst{1}\fnsep\thanks{\email{knechtli@physik.uni-wuppertal.de}}
}

\institute{
  Department of Physics, Bergische Universit{\"a}t Wuppertal, 
              Gaussstr.~20, 42119~Wuppertal, Germany
          }

\abstract{%
  We review selected lattice results on the charmonium spectrum and first attempts to
search for the existence of exotic states. The hadro-quarkonium model was proposed
to interpret some of the exotic states as a quarkonium core inside a hadron. We present a
lattice study of the hadro-quarkonium model in the limit of static quarks. The charm
quark decouples in low energy observables and binding energies of charmonium. In a model
calculation we are able to evaluate the corrections to decoupling of the charm quark
in the continuum.
}
\maketitle
\section{Introduction}
\label{intro}

In the period 1974-1977, 10 charmonium $c\bar{c}$ resonances were discovered. None were discovered in 1978-2001.
Since 2002 new $c\bar{c}$'s were found by BaBar, Belle, CLEO-c, CDF, D0 \cite{Olsen:2015zcy}.
Some of these states are exotic (or charmoniumlike) and are labeled as $X$, $Y$ or $Z$ candidates.
There are several criteria to classify states as exotic based on quantum numbers, electric charge,
supernumerary states, decays etc. \cite{Lebed:2016hpi}.

Charmonium resonances can be studied on the lattice. The masses are well understood if the states are treated as stable.
States above the open charm thresholds $D\bar{D}$ etc. decay strongly and multi-hadron channels need to be included.
On a finite Euclidean lattice there is no dynamical real-time and no asymptotic states.
A workaround is that scattering data can be inferred from the spectrum of QCD in a finite volume below the inelastic threshold
\cite{Luscher:1986pf,Luscher:1990ux}. For a recent review
on scattering on the lattice see \cite{Briceno:2017max}.

\section{Charmonium and exotics}
\label{sec-1}

\subsection{Charmonium and its excited states}
\label{sec-1-charmonium}

\begin{figure}[t]
\centering
\includegraphics[width=0.8\textwidth,clip]{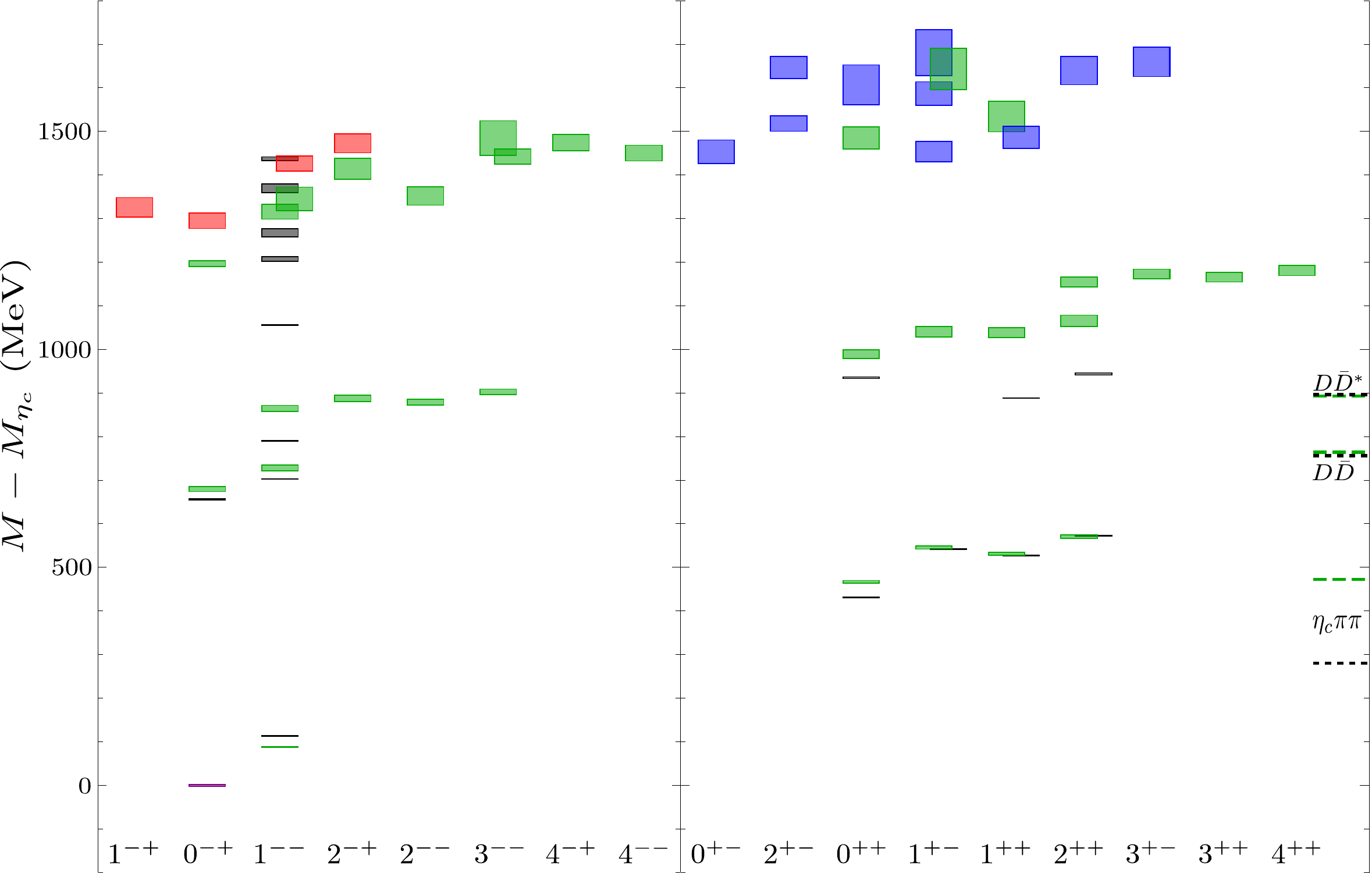}
\caption{Charmonium spectrum from \cite{Cheung:2016bym}.
  Results on the $M_\pi=236\mev$ ensemble are compared to PDG (black).
  Many states (green) follow the $n^{2S+1}L_J$ pattern of quark potential models.
  Excess states (red) and (blue) are also identified. They are consistent with
  being hybrid mesons coupled to a gluonic excitation $\bar{c}cg$.
  Four of the hybrids have exotic quantum numbers $J^{PC}=0^{+-}$, $1^{-+}$ and $2^{+-}$.
}
\label{f:charmonium}       
\end{figure}

In figure~\ref{f:charmonium} we show the results of the charmonium spectrum from
a calculation by the Hadron Spectrum Collaboration \cite{Liu:2012ze,Cheung:2016bym}.
They use $128\times 24^3$ and $256\times 32^3$ ensembles generated with
$\nf=2+1$ dynamical quarks. The mass of the strange quark is
$m_\mathrm{strange}\approx m_\mathrm{strange}^{\mathrm{phys}}$,
the mass of the light quarks is  $m_\mathrm{up}=m_\mathrm{down}=m_\mathrm{light}$
and corresponds to $M_\pi=391\mev$ ($128\times 24^3$) and
$M_\pi=236\mev$ ($256\times 32^3$).
The charm quark is quenched (only valence, not dynamical),
its mass is tuned to reproduce the physical $\eta_c$ mass.
The lattices are anisotropic, the spatial lattice spacing is
$a_s\approx0.12\fm$ and the temporal lattice spacing is considerably smaller
$a_t\approx0.034\fm$. The charmonium spectrum is extracted from
two-point functions of charm quark bilinears. No multi-hadron operators are included.
States above threshold are treated as stable, which means that their mass
is accurate up to the hadronic width.
Improved techniques, such as
operator construction \cite{Dudek:2007wv,Dudek:2009qf,Dudek:2010wm},
distillation \cite{Peardon:2009gh},
variational method \cite{Michael:1985ne,Luscher:1990ck} are used.
Charm-annihilation (disconnected) diagrams are not included.
They are expected to  be small since they are suppressed by the
OZI (Okubo, Zweig, Iizuka) rule, cf. \cite{Levkova:2010ft}.
Another calculation of the charmonium spectrum can be found in
\cite{Kalinowski:2015bwa}.

\subsection{The $X(3872)$}
\label{sec-1-X}

\begin{figure}[t]
  \centering
  \sidecaption
\includegraphics[width=0.5\textwidth,clip]{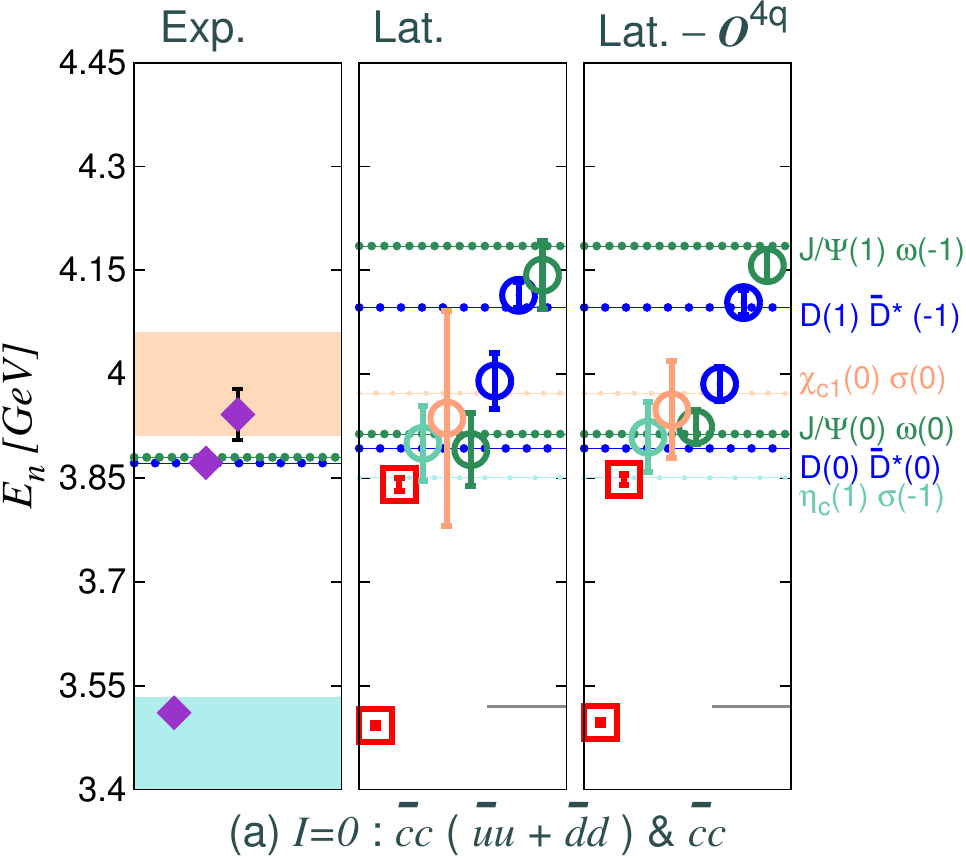}
\caption{Study of the $X(3872)$ on the lattice \cite{Padmanath:2015era}.
  The basis of lattice operators include $\bar{c}c$, two-meson and
  tetraquark $O^{4q}$ operators.
}
\label{f:x3872}
\end{figure}

The first charmoniumlike state which was experimentally observed
\cite{Choi:2003ue} is the $X(3872)$. Its quantum numbers are
$J^{PC}=1^{++}$ and its mass is $M=3871.69\pm0.17\mev$, which is
equal to the sum of the $D^0$ and $D^{*0}$ masses,
$m_{D^0}+m_{D^{*0}}=3871.69\pm0.09\mev$ \cite{Olsen:2015zcy}.
On the lattice it has been studied in \cite{Padmanath:2015era}
on $\nf=2$ lattices with a pion mass $m_\pi=266\mev$,
lattice spacing $a=0.12\fm$
and lattice size $L\simeq2\fm$.
It appears as an additional energy eigenstate compared to
the non-interacting spectrum of two meson states, see the $n=2$ level
(red square) in the left plot of figure~\ref{f:x3872}.
The interpretation of $X(3872)$ as a pure molecule or a pure tetraquark
is unlikely. Leaving out the tetraquark operators (right hand block) does not affect much the spectrum. Tetraquark operators are studied in \cite{Cheung:2017tnt}.

\section{Hadro-charmonium}
\label{sec-2}

\begin{figure}[t]
\centering
\includegraphics[width=0.7\textwidth,clip]{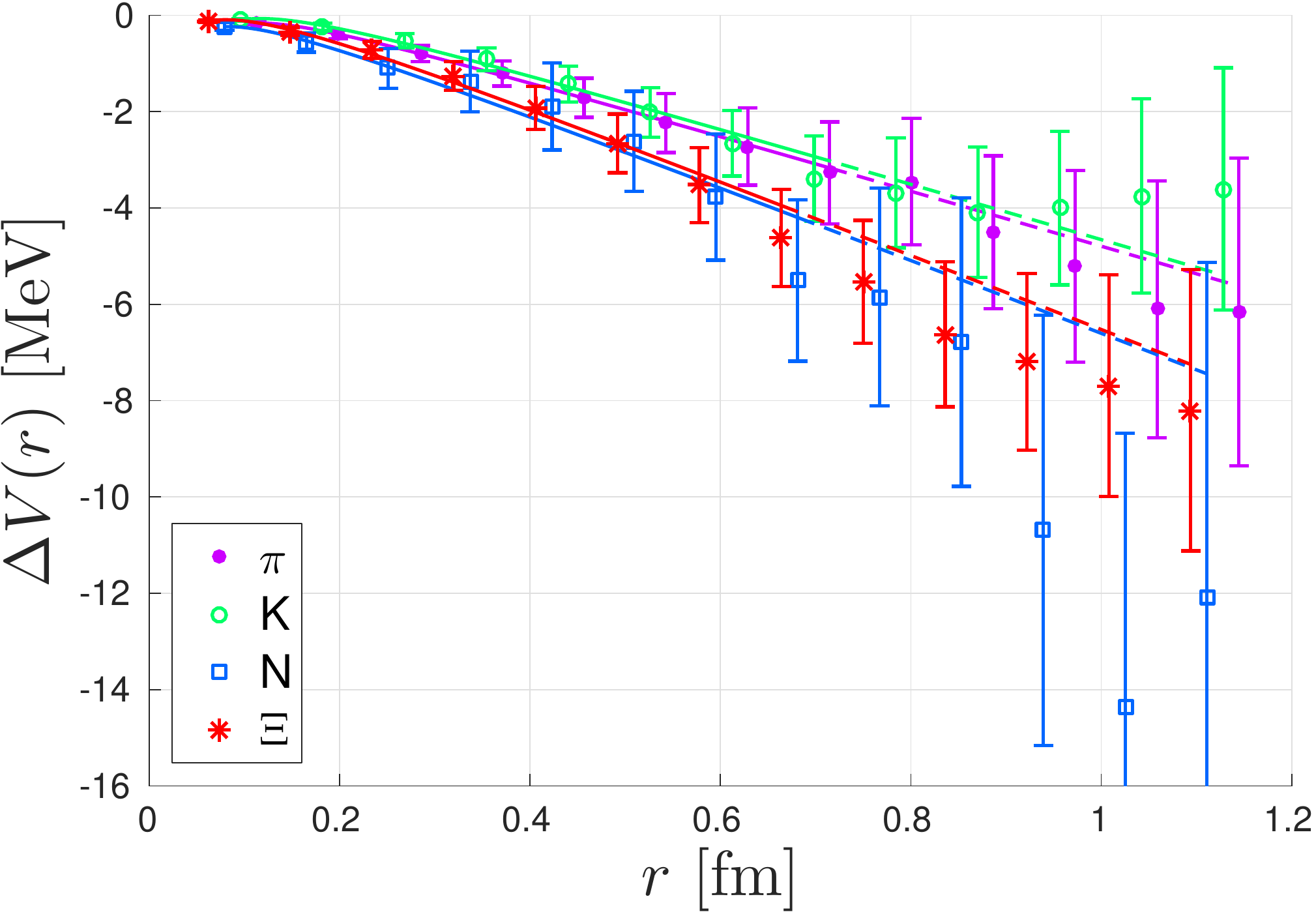}
\caption{The modification $\Delta V_H(r)$ of the static potential due to the
  presence of a hadron \cite{Alberti:2016dru}.
  The curves represent the parametrization
$\Delta V_H(r) = \Delta\mu_H - \Delta c_H / r + \Delta\sigma_H r$.
  The parameters $\Delta\mu_H$, $\Delta c_H$ and $\Delta\sigma_H$
describe the modifications to the Cornell potential.
}
\label{f:hadroqua}
\end{figure}

LHCb found two pentaquark candidates $P_{c}^{+}$ of exotic
quark content $uudc\bar{c}$ in the decay
$\Lambda_b\rightarrow (J/\psi~p)~K$
\cite{Aaij:2015tga,Aaij:2016phn}.
Systems consisting of 5 quarks (4 $q$, 1 $\bar{q}$) are very difficult to
study directly on the lattice, in particular if many decay channels are possible.
A $20\mev$ binding energy was reported for the charmonium-nucleon system for a
rather large light quark mass ($m_\pi\approx800\mev$) and coarse lattice spacing
$a\approx0.145\fm$ in \cite{Beane:2014sda}.

A possible explanation of such exotic penta-quark states with a $c\bar{c}$
content is the hadro-quarkonium model. The idea is that of a
quarkonium core embedded in a light hadron cloud \cite{Dubynskiy:2008mq} and is
based on an attractive color dipole-dipole van der Waals interaction
between the point-like quarkonium and the hadron.
The LHCb pentaquark candidates could correspond in this model to the following
close-by charmonium-baryon systems:
\begin{eqnarray}
  J^P=\frac{3}{2}^{-}: & m(\Delta)+m(J/\psi)\approx4329\mev & \mbox{vs.}\;
  P_{c}^{+}(4380)\; \mbox{(width $200\mev$)} \nonumber \\
  J^P=\frac{5}{2}^{+}: & m(N)+m(\chi_{c2})\approx4496\mev &  \mbox{vs.}\;
  P_{c}^{+}(4450)\; \mbox{(width $40\mev$)} \nonumber
\end{eqnarray}
In \cite{Alberti:2016dru} we performed a lattice study of the hadro-quarkonium
in the static approximation $m_Q\to\infty$ for the quarkonium $\bar{Q} Q$.
In this limit the interaction energy of quarkonium is given by the
static quark potential $V(r)$ ($r$ is the separation of the
static quarks) and can be calculated from lattice simulations, cf.
\cite{Knechtli:2017muu}.
What needs to be answered if the following question.
Does the static potential become more or less attractive,
when a light hadron $H$ is ``added''?
The shift of the static potential $\Delta V_H$ can be computed from
a suitable ratio of correlators and we refer to \cite{Alberti:2016dru}
for the details.
In figure~\ref{f:hadroqua} we show the results
for the pion, kaon, nucleon and $\Xi$.
They were obtained by
analyzing the $\nf=2+1$ ensemble ``C101'' ($96\times 48^3$ sites, 1552 gauge
configurations) generated by
the Coordinated Lattice Simulations (CLS) consortium \cite{Bruno:2014jqa}.
The pion mass is
$\mpi=220\mev$,
the kaon mass is $\mK=470\mev$,
the lattice volume is
$L=4.1\fm$ and the lattice spacing is
$a=0.0864(11)\fm$ \cite{Bruno:2016plf}.
In order to be able to measure $\Delta V_H$ a high statistics was necessary.
Notice that we only show distances $r$ smaller
than the string breaking distance $r_c\approx 1.2\fm$ \cite{Bulava:2019iut,Bali:2005fu}.

\begin{figure}[t]
\centering
\sidecaption
\includegraphics[width=0.5\textwidth,clip]{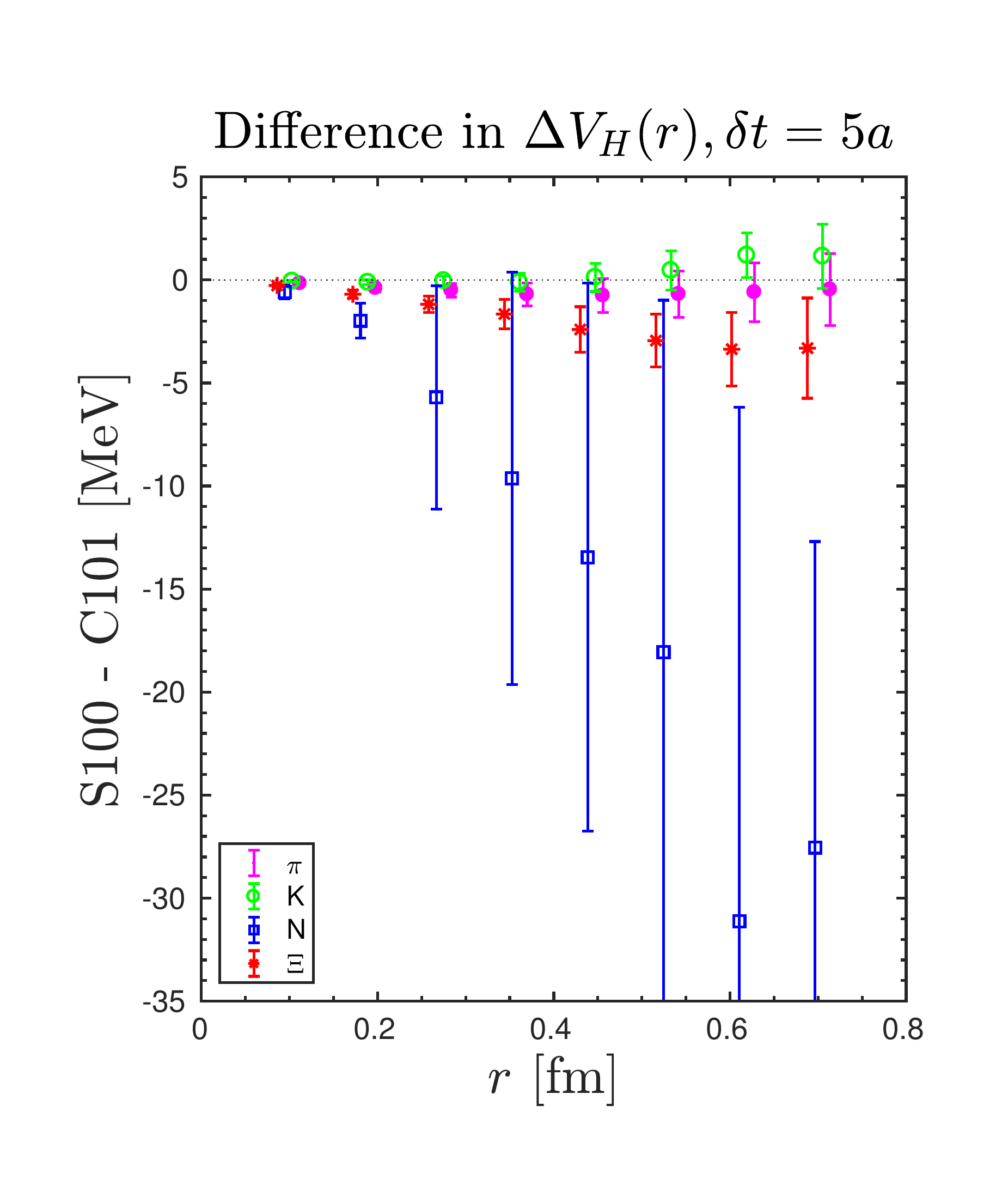}
\caption{Volume study of the shift of the static potential $\Delta V_H$
  in presence of various hadrons. We plot the difference
  of the shifts between lattices of size $L=2.8\fm$ (``S100'') and $L=4.1\fm$ (``C101'').
}
\label{f:hadroqua_volume}
\end{figure}

A potential source of systematic effects is the finite size $L$ of the lattice.
We therefore performed the calculation also on the
CLS ensemble ``S100'' ($128\times32^3$) which has the
same lattice spacing $a$ and quark masses as ``C101'' but
smaller size $L=2.8\fm$.
The statistics is $940$ configurations times 10 hadron sources
(forward and backward propagating). In order to
check for finite volume effects we compute the difference
$
    \Delta V_H(r)^{{\rm S100}}-\Delta V_H(r)^{{\rm C101}}
$
which is shown in figure~\ref{f:hadroqua_volume}. We do not observe
significant finite volume effects for distances $r>0.3\fm$.
Only the statistical errors are shown.  

The phenomenological implications of the potential shift $\Delta V_H$
can be derived in a non-relativistic approach (potential NRQCD) to describe
the quarkonia $\overline{Q}Q$. There the quarkonium levels are obtained by
solving the Schr\"odinger equation. Repeating the calculation with
the static potential in the vacuum $V_0$ and with the modified potential
$V_H=V_0+\Delta V_H$ yields the changes in the quarkonium levels.
For charmonium
we find that the energies decrease, indeed indicating
that charmonium ``inside'' a hadron $H$ is energetically favorable.
But the size of the shifts is only of few $\mev$'s \cite{Alberti:2016dru}.
One should keep in mind though that
relativistic corrections are not small for charmonium and the
mass of the hosting baryons is comparable to that of the charm quark.

\section{Decoupling of the charm quark}
\label{sec-3}

At present most simulations of lattice QCD are done with $\nf=2+1$ dynamical
quarks (up, down and strange). The inclusion of dynamical heavy quarks (charm)
requires
\begin{itemize}
  \item high precision in low energy observables to resolve tiny charm-quark loop effects
  \item small lattice spacings to control cut-off effects proportional to the heavy quark mass
\end{itemize}
In the following we want to illustrate these points.

\subsection{Test of the low energy effective theory}
\label{sec-3-lowenergy}

A heavy quark of mass $M$ decouples from physical processes at energies
$E\ll M$. The latter can be described in terms of an effective theory which
arises from integrating out the heavy quark \cite{Weinberg:1980wa}.
The effective theory depends on the heavy quark mass $M$
\begin{enumerate}
  \item through the gauge coupling and the light fermion masses and
  \item through higher dimensional operators multiplied by inverse powers of $M$.
\end{enumerate}
In perturbation theory the power effects can be separated. For sufficiently
heavy quark masses $M$ the matching relation between the gauge coupling of the
effective theory (without the heavy quark) and the
gauge coupling of the fundamental
theory (with the heavy quark) can be computed using perturbation theory
\cite{Weinberg:1980wa,thresh:BeWe,Chetyrkin:2005ia,Grozin:2011nk}.
The perturbative matching relations have been used for the charm quark and the
bottom quark by the ALPHA collaboration to obtain
$\alpha_{\overline{\text{MS}}}(M_Z)   = 0.1185(8)(3)$ from simulations of
QCD with $\nf=2+1$ quarks \cite{Bruno:2017gxd}. In particular the
question of applicability of perturbation theory for the charm quark arises.
While perturbation theory by itself converges very well, a check of
non-perturbative effects in the matching and of the size of power
corrections is desirable.
\begin{figure}[t]
\sidecaption
\centering
\includegraphics[width=0.5\textwidth,clip]{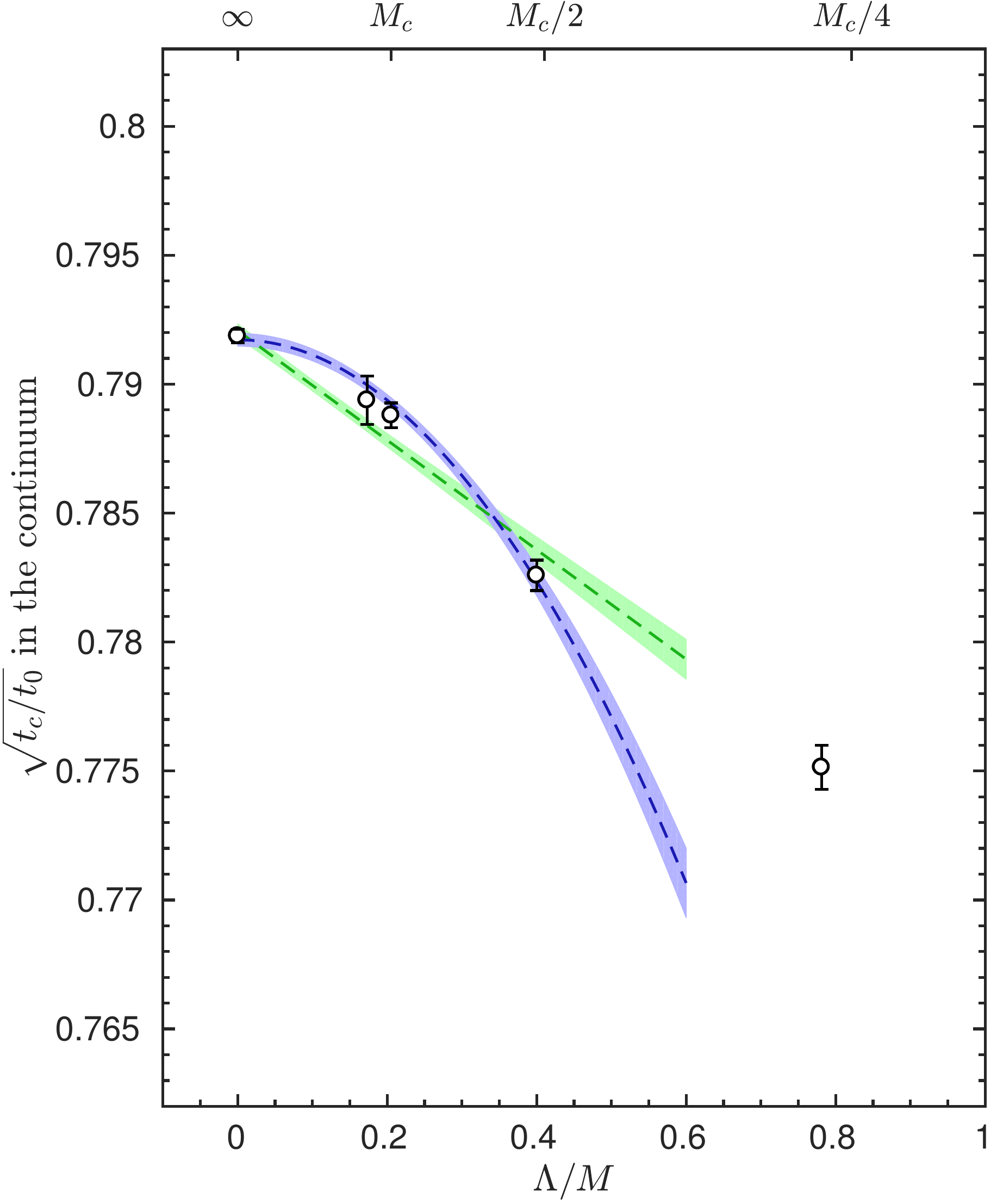}
\caption{The ratio of scales in eq.~\ref{e:ratio} computed in QCD with
  $\Nf=2$ heavy quarks of mass $M$ and in the Yang--Mills theory corresponding
  to $M=\infty$. The dashed line in the blue band is a fit to the leading behavior in
  eq.~\ref{e:ratio} with fit parameter $k$. It is compared to a fit linear
  in $1/M$ (dashed line in the green band). From \cite{Knechtli:2017pxe}.
}
\label{f:powcorr}
\end{figure}

With this motivation we studied a model on the lattice \cite{Bruno:2014ufa}.
We simulated QCD with $\Nf=2$ degenerate quarks of mass
$1.2\,M_\mathrm{charm} \gtrsim M \gtrsim M_\mathrm{charm}/8$
and compared to pure Yang--Mills (YM) theory, which is the leading order
in the effective theory at low energy.
We can afford very small lattice spacings down to $a=0.023\fm$ and
control the continuum limit.
We computed in both theories low energy hadronic scales.
In particular scales based on the gradient flow \cite{flow:ML,flow:Herbert} can
be measured very precisely in lattice simulations.
For example we consider the ratio of the scale $\sqrt{t_0}$~\cite{flow:ML}
and its cousin $\sqrt{t_c}$, cf. \cite{Athenodorou:2018wpk}.
For the ratio of the two scales the effective theory predicts \cite{Knechtli:2017xgy}
\begin{equation}\label{e:ratio}
    \left.\sqrt{ t_c(M) / t_0(M) }\right|_{\Nf=2}
    = \left.\sqrt{ t_c / t_0 }\right|_\mathrm{YM}
    +  k / M^2 \,,
\end{equation}
with leading corrections proportional to $1/M^2$.
Figure~\ref{f:powcorr} shows the results for the ratio in the
continuum limit. It can be well fitted by eq.~\ref{e:ratio}
(dashed line in the blue band) down to masses
of about $M_\mathrm{charm}/2$.

The $1/M_\mathrm{charm}^2$ corrections in the ratio eq.~\ref{e:ratio}
originate from charm-quark loops and are found to be
very small, 0.4\% when decoupling two charm quarks like in figure~\ref{f:powcorr}.
Another class of effects of charm loops at low energy is the connection
of the fundamental scales (the $\Lambda$-parameters) of the theory with and
without the charm quark. They have been calculated in \cite{Athenodorou:2018wpk}
in the same model as just described. From this calculation we conclude that
non-perturbative effects in the conversion of the three-flavor and four-flavor
Lambda parameters can be neglected at a level down to 1\% accuracy.
Notice that the precision for the $\Lambda$-parameter is currently at the level of around
4\% \cite{Aoki:2016frl,Olive:2016xmw}.

\subsection{Charm loop effects in charmonium}
\label{sec-3-charmonium}
\begin{figure}[t]
\centering
\includegraphics[width=0.8\textwidth,clip]{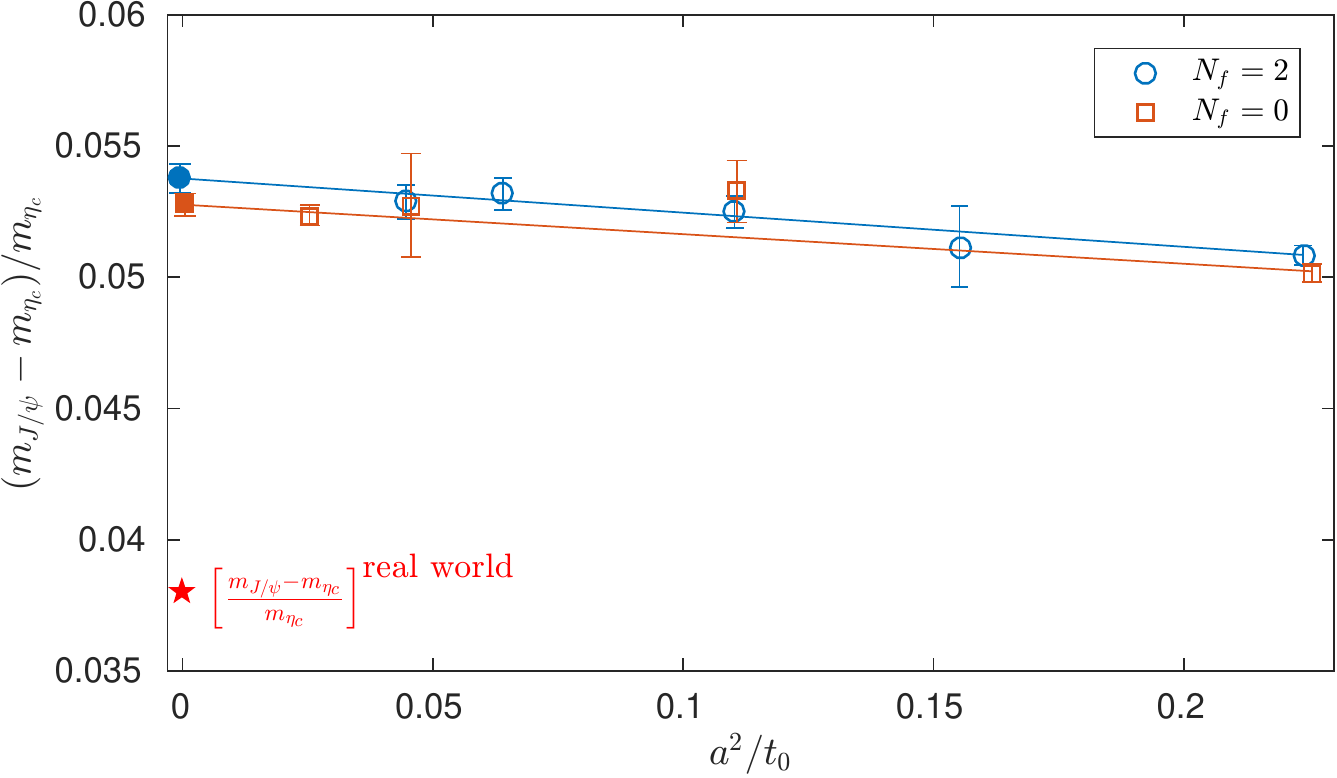}
\caption{Continuum extrapolation of the hyperfine splitting in $\nf=2$ and $\nf=0$ QCD \cite{charmloopeffects}.
}
\label{f:hyperfine}
\end{figure}

We consider now quantities which have an explicit valence charm quark
and want to compute the impact of the dynamical charm quark through loops.
This we do in our model study by comparing
$\Nf=0$ QCD (Yang--Mills) and QCD with $\Nf=2$ degenerate charm quarks
\cite{Korzec:2016eko,Cali:2018owe}.
The two theories have to be matched and for this we use
decoupling for the scale
$\left[\sqrt{t_0(M_\mathrm{charm})}\right]^{\Nf=2}=\left[\sqrt{t_0}\right]^{\Nf=0}$.
The charm-quark mass $M_\mathrm{charm}$ is then fixed by requiring
$\sqrt{t_0}m_{\eta_c}\equiv 1.8075$ in both theories.
This corresponds approximately to the physical $m_{\eta_c}$
In this setting differences in the hyperfine splitting
\begin{equation}\label{e:hyperfine}
  [(m_{J/\psi}-m_{\eta_c})/m_{\eta_c}]^{\nf=2} - [(m_{J/\psi}-m_{\eta_c})/m_{\eta_c}]^{\nf=0}
\end{equation} 
are due to charm loop effects.
Large cut-off effects have been previously observed in the hyperfine splitting
$m_{J/\psi}-m_{\eta_c}$ \cite{Cho:2015ffa,Rae:2015fwa}. Therefore fine lattices
are needed, which we can afford in our model study.

Figure~\ref{f:hyperfine} shows the continuum extrapolations of the hyperfine
splittings in eq.~\ref{e:hyperfine}. The relative difference between $\nf=2$ (filled blue circle)
and $\nf=0$ (filled red square) is $0.018(13)$. The discrepancy with the
physical value (red star) is probably due to neglected effects of light sea quarks, disconnected
contributions and electromagnetism.

\section{Conclusions}
\label{concl}

Lattice QCD provides the techniques to study charmonium resonances.
Candidates for exotic states can be identified through lattice simulations
and their nature may be elucidated.

The hadro-charmonium has been tested in the static limit.
It yields stronger bindings
of charmonium ``inside'' hadrons but only by few $\mev$'s like deuterium binding.
The modification of the static potential ``inside'' a hadron is also interesting
for charmonium in medium.

Decoupling of the charm quark at low energies is consistent with the effective
theory beyond leading order.
Decoupling also applies to differences of binding energies of charmonium. The relative
effects of charm loops in the hyperfine splitting are below 2\% for
a model study of QCD with two charm quarks.

{\bf Acknowledgments.}
I thank the organizers for the nice conference in Novosibirsk.
I acknowledge financial support from
the European Union's Horizon 2020 research and innovation programme under the Marie Sklodowska-Curie grant agreement Number 642069.

%
\bibliography{charm}
%
%
%
%

\end{document}